\documentclass[preprint,onecolumn,showkeys]{revtex4-1}
\usepackage{amssymb}
\usepackage{amsmath}
\usepackage{graphicx}
\usepackage{subfigure}
\usepackage{upgreek}
\begin{document}

\title{Strong and Tunable Spin--Orbit Coupling of One-Dimensional
  Holes in Ge/Si Core/Shell Nanowires}

\author{Xiao-Jie Hao$^{(1,2)}$}
\author{Tao Tu$^{(1)}$}
\author{Gang Cao$^{(1)}$}
\author{Cheng Zhou$^{(1)}$}
\author{Hai-Ou Li$^{(1)}$}
\author{Guang-Can Guo$^{(1)}$}
\author{Wayne Y. Fung$^{(2)}$}
\author{Zhongqing Ji$^{(2)}$}

\author{Guo-Ping Guo$^{(1)}$}
\email{gpguo@ustc.edu.cn}

\author{Wei Lu$^{(2)}$}
\email{wluee@eecs.umich.edu}

\affiliation{$^{(1)}$Key Laboratory of Quantum Information, University
  of Science and Technology of China, Chinese Academy of Sciences,
  Hefei 230026, People's Republic of China\\
  $^{(2)}$Department of Electrical Engineering and Computer Science,
  The University of Michigan, Ann Arbor, Michigan 48109, USA}

\date{\today }

\begin{abstract}
  We investigate the low-temperature magneto-transport properties of
  individual Ge/Si core/shell nanowires. Negative magneto-conductance
  was observed, which is a signature of one-dimensional weak
  antilocalization of holes in the presence of strong spin--orbit
  coupling. The temperature and back gate dependences of phase
  coherence length, spin--orbit relaxation time, and background
  conductance were studied. Specifically, we show the spin--orbit
  coupling strength can be modulated by more than five folds with an
  external electric field. These results suggest the Ge/Si nanowire
  system possesses strong and tunable spin--orbit interactions and may
  serve as a candidate for spintronics applications.
\end{abstract}
\keywords{One-dimension, spintronics, nanowire, weak antilocalization,
  magneto-conductance}

\maketitle
%%%%%%%%%%%%%%%%%%%%%%%%%%%%%%%%%main txt%%%%%%%%%%%%%%%%%%%%%%%%%%%%%%%%%%%%

Semiconductor nanowires exhibit novel electrical, optical, and
mechanical properties and offer substantial potential as building
blocks of nanodevices owing to their one-dimensional structure \cite
{Lieber2003,Lu2006,Samuelson2006}. Besides being a good candidate for
high performance electronic devices, nanowires may also be used in the
field of spintronics, which involves exploration of the extra degrees
of freedom provided by electron spin, in addition to those due to
electron charge \cite{Wolf2001,Awschalom2002}. In particular, Ge/Si
core/shell nanowires represent a unique one-dimensional system for
exploring quantum coherence phenomenon at the nanoscale because of
their high hole mobility and strong quantum confinement effects \cite
{Lu2005,Xiang2006,Xiang2006Josephson,Liang2007}. In addition, as
compared with III--V materials (for example, InAs) where hyperfine
coupling limits the electron spin coherence, the prospect of long
coherence times in group IV materials due to the predominance of
spin-zero nuclei has stimulated several proposals and significant
experimental effort for spin-based quantum information applications
\cite{Awschalom2002,Hu2007,Roddaro2008}. One key to realizing such
promise is the utilization of spin--orbit interaction which can be
controlled by tuning the applied gate voltages. However there are few
investigations of spin--orbit interactions and corresponding
relaxation times in group IV semiconducting nanowires. Here, we report
on low-temperature magneto-conductance measurements of individual
gated Ge/Si core/shell nanowires in transverse magnetic field. The
observed negative magneto-conductance data are consistent with the
one-dimensional weak antilocalization effect of holes in the presence
of strong spin--orbit interactions
\cite{Roddaro2008,Hikami,Beenakker,Kurdak,Hansen}. We extract the
phase coherence length/time, the spin--orbit coherence length/time,
and the spin--orbit coupling constant, and verify both the hole phase
coherence and the spin--orbit coupling can be tuned in our system as a
function of gate voltage. The gate-tunable spin--orbit coupling
strength suggests the Ge/Si core/shell nanowire is a candidate
platform for designing future spintronics applications
\cite{Awschalom2002}.

\begin{figure}[tbp]
\centerline{\subfigure[$I_{ds}-V_{ds}$ characteristic]
    {\label{fig:IV}\includegraphics[width=0.45\columnwidth]{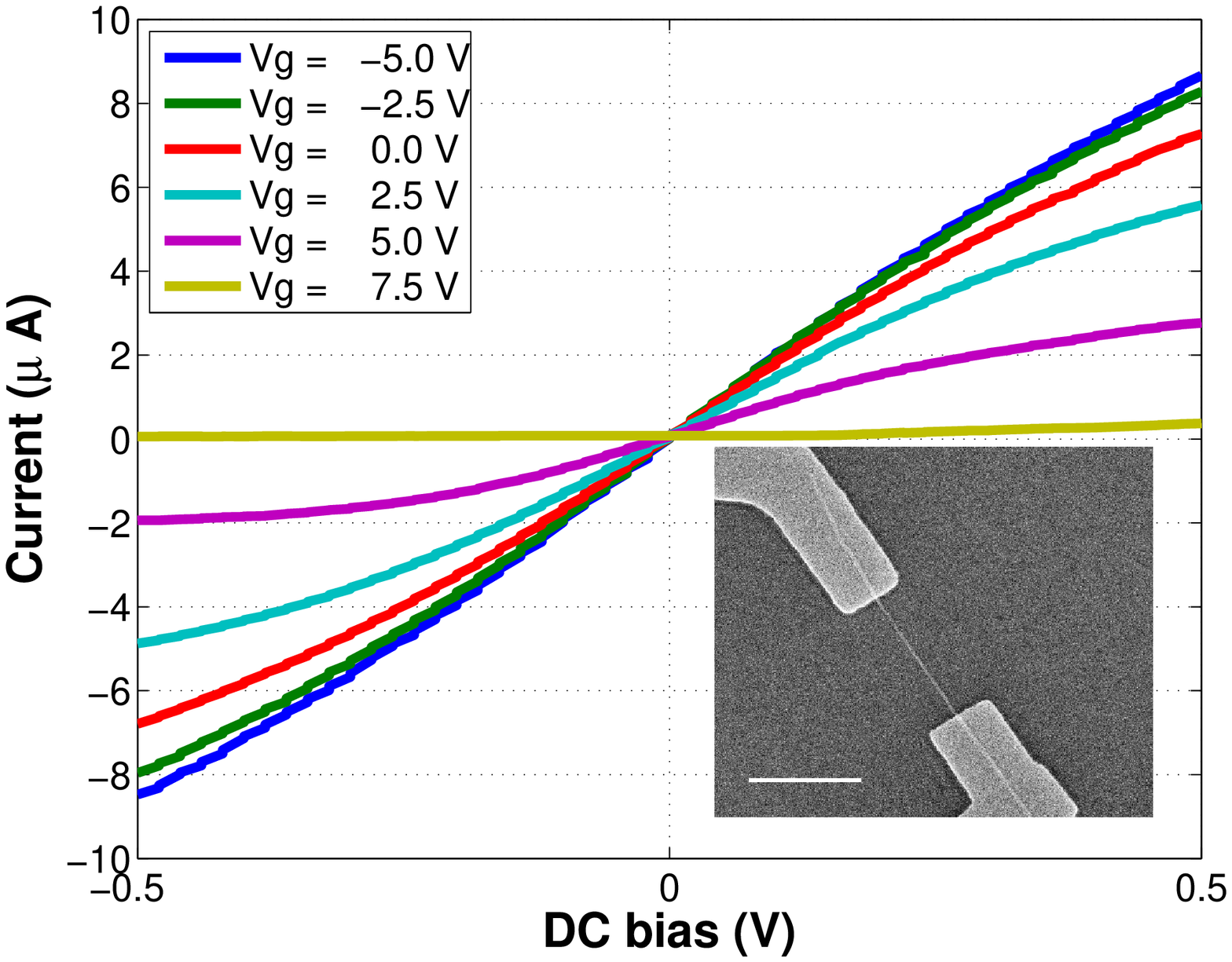}}
    \subfigure[Magneto-conductance]
    {\label{fig:GB}\includegraphics[width=0.45\columnwidth]{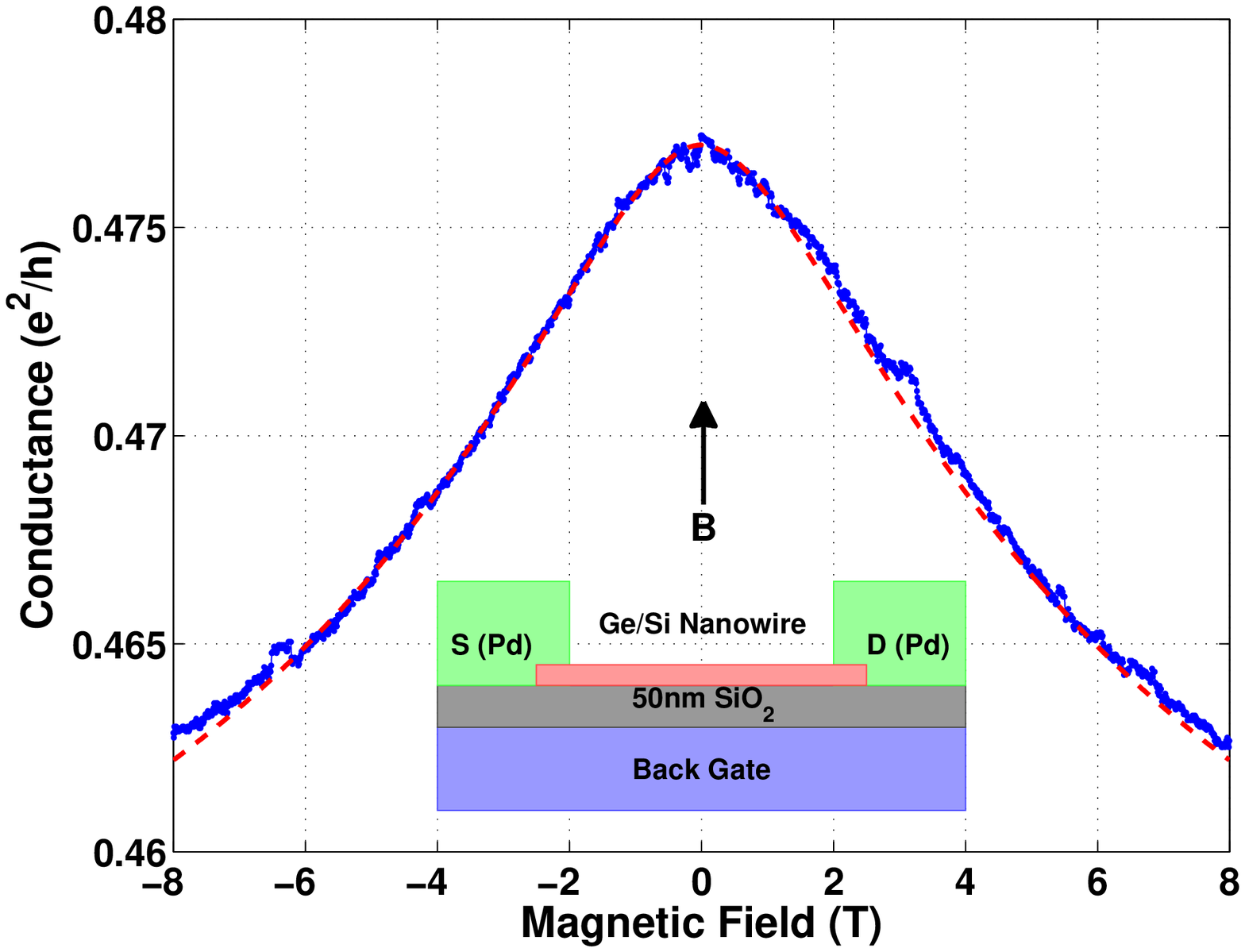}}}
  \caption{\Subref{fig:IV} Current through the device as a function of
    bias voltage at different back gate voltages at $400$ mK. (inset)
    SEM image of a similar device (scale bar: $1$ $\upmu$m).
    \Subref{fig:GB} The measured conductance (blue dotted line) as a
    function of magnetic field at back gate voltage of $-1$ V,
    together with the fitting curve (red dashed line). (inset)
    Schematics of the experiment setup. }
\label{fig:conductance}
\end{figure}

\begin{figure}[tbh]
\centering{\includegraphics[width=0.55\columnwidth]{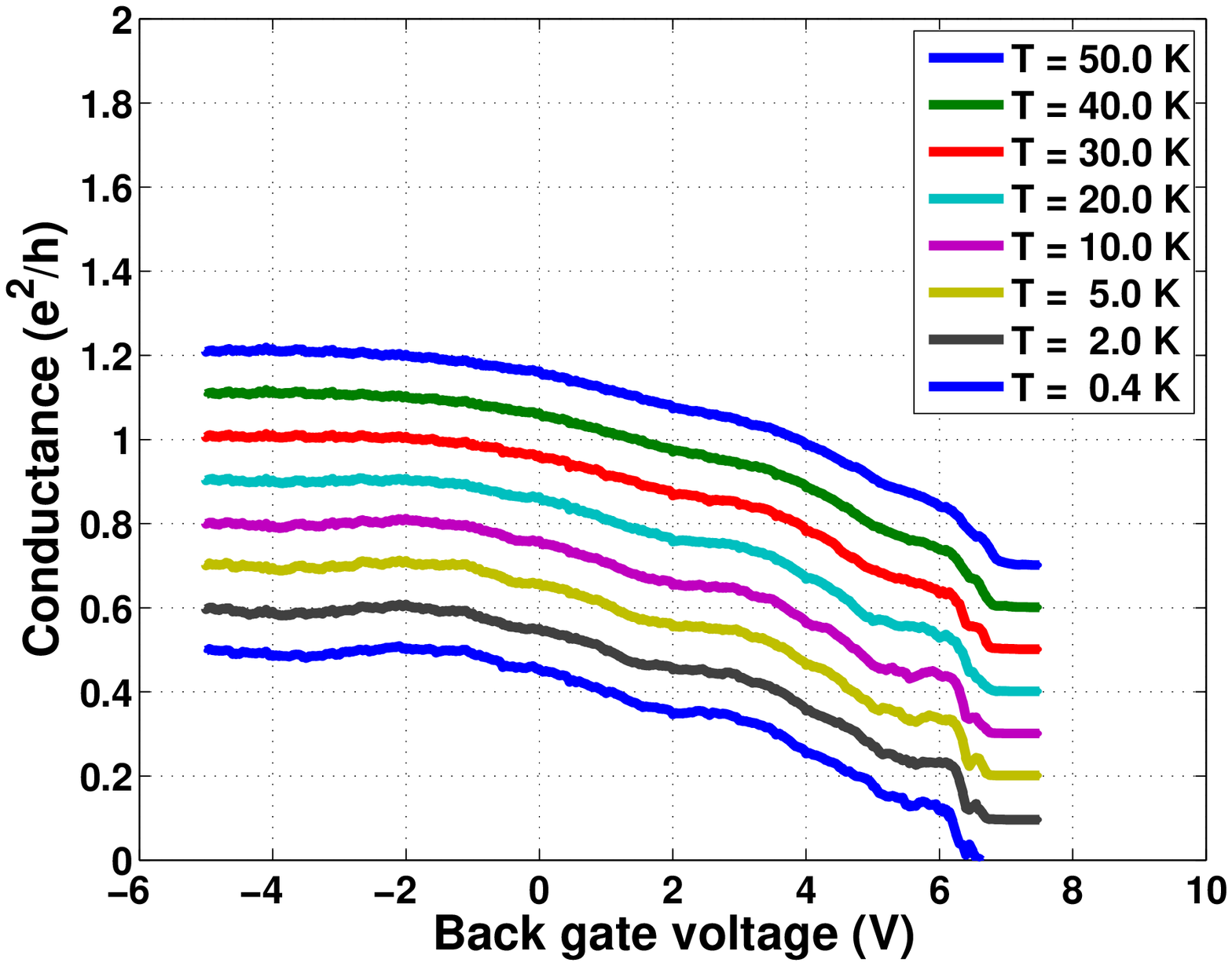}}
\caption{Conductance as a function of gate voltage at different
  temperatures at a source drain bias of $V_{ds}=20$ mV. The $n$th
  curve from bottom to top is shifted by $0.1\times (n-1)\times
  e^{2}/h$ vertically for clarity.}
\label{fig:GVgT}
\end{figure}

\begin{figure}[tbh]
\centering{\includegraphics[width=0.55\columnwidth]{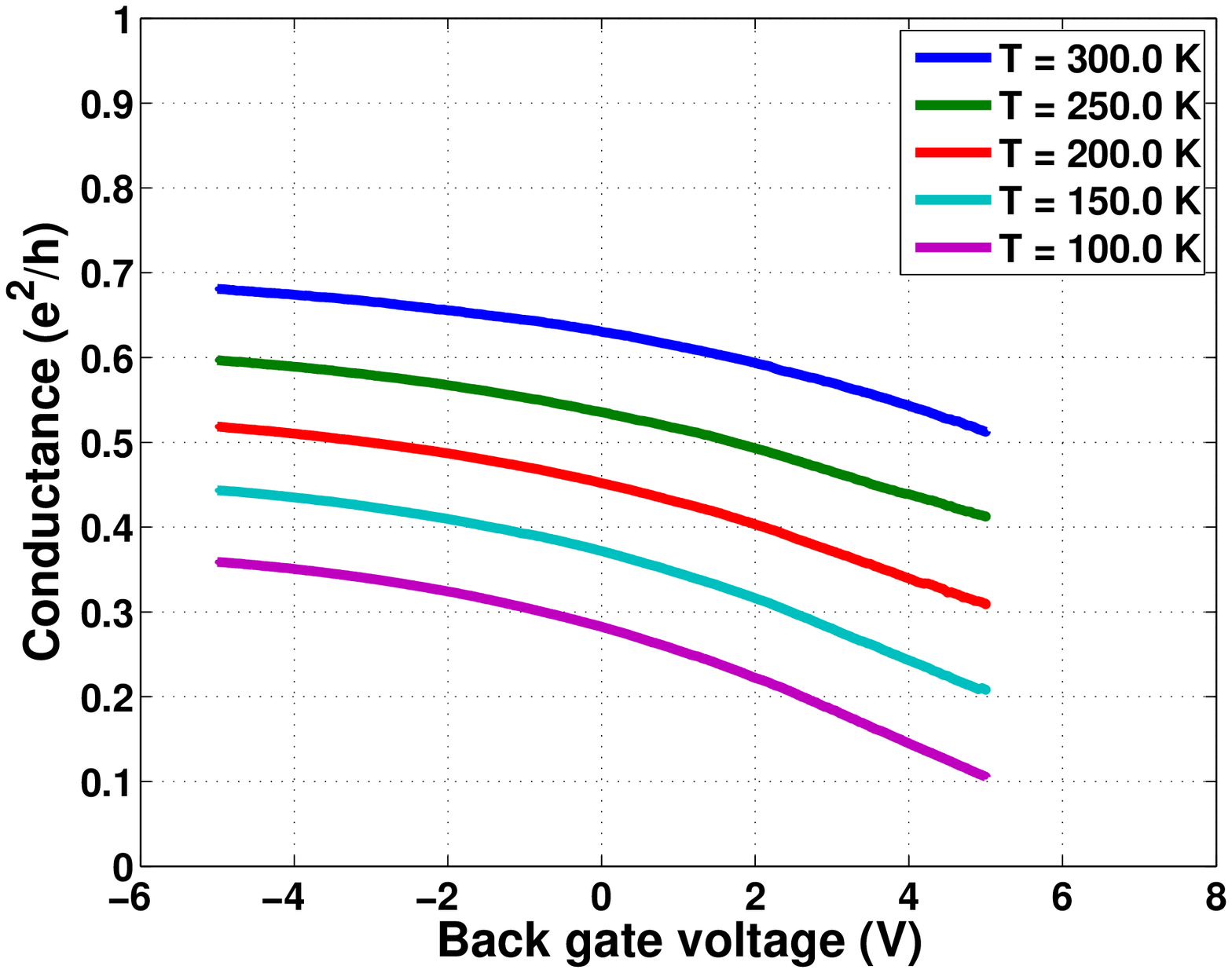}}
\caption{Conductance as a function of gate voltage at different
  temperatures at a source drain bias of $V_{ds}=1$ V. The $n$th curve
  from bottom to top is shifted by $0.1\times (n-1)\times e^{2}/h$
  vertically for clarity.}
\label{fig:GVgT2}
\end{figure}

The undoped Ge/Si core/shell nanowires studied here, which have an
average germanium core of $10$ nm and silicon shell of $2$ nm, were
grown using a two-step chemical vapor deposition process reported
earlier \cite{Lu2005}. The nanowires were suspended by ultrasound
sonication in isopropyl alcohol, then deposited onto $50$ nm thick
oxide on a degenerately doped Si substrate. The Si substrate was also
used as a bottom gate electrode. After nanowire transfer, source and
drain contacts were defined by electron-beam lithography. To make good
contacts between metal leads and the Ge core of the nanowires, the
samples were treated in buffered hydrofluoric acid for $3$ s to remove
any native oxide on top of the Si shell before electron-beam
evaporation of source/drain electrodes (palladium, $50$ nm thick). A
total of three devices from the same batch of nanowires were studied,
and all have produced remarkably remarkably similar results. For
consistency, we present here the data (including the mobility and
density value) from only one device. The scanning electron microscope
(SEM) image of a similar device is shown in the inset of
Fig.~\ref{fig:IV}. In Fig.~\ref {fig:IV}, the current is plotted
against bias voltage at several different back gate voltages at the
temperature of $400$ mK. The low-temperature linearities of these
$I_{ds}-V_{ds}$ characteristics, especially at zero bias, verify good
Ohmic contacts between metal leads and the nanowire were achieved. We
also observed signs of possible multiple subbands transport through
the nanowire at different temperatures (see Supporting Information
Fig.~\ref{fig:GVgT}) similar to our earlier studies \cite{Lu2005}.
Using the transport data and taking into account of the gate
capacitance $C_{g}$, which is given by the cylinder-on-plane model
\cite {Ramo}, the hole mobilities of the device can be calculated as
followsing \cite {Lu2005,Dai2003}:

\begin{equation}
  \mu = \frac{dI_{ds}}{dV_{g}}\times \frac{L^{2}}{C_{g}}\times
  \frac{1}{V_{ds}} \label{eq:mu}
\end{equation}
Here, $dI_{ds}/dV_{g}$ was obtained from the linear region of the
$I_{ds}-V_{g}$ data (see Supporting Information Fig.~\ref{fig:GVgT}
and Fig.~\ref{fig:GVgT2}), $L=1.5$ $\upmu$m is the length of the
nanowire between two contacts of the device, and $V_{ds}$ is the
source-drain voltage at which the $I_{ds}-V_{g}$ data were taken. A
mobility of $\sim 300$ cm$^{2}$V$^{-1}$s$^{-1}$ at room temperature
and of $\sim 600$ cm$^{2}$V$^{-1}$s $^{-1}$ at liquid Helium
temperature were determined. These values are consistent with previous
results \cite{Xiang2006} in Ge/Si core/shell nanowires.

\begin{figure}[tbh]
\centering{\includegraphics[width=0.55\columnwidth]{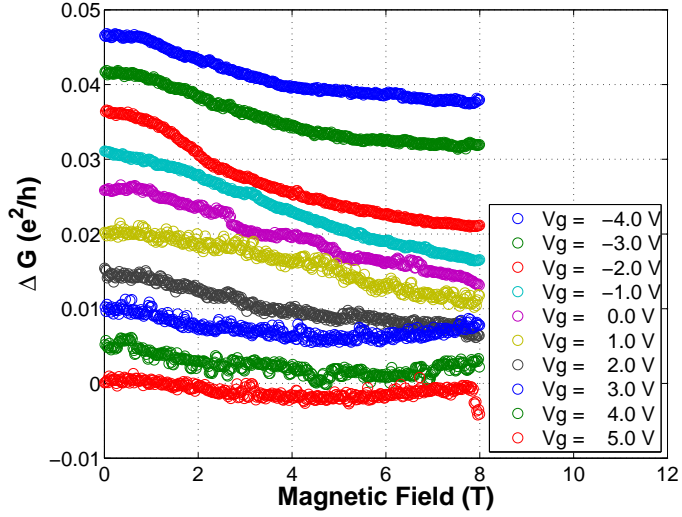}}
\caption{Magneto-conductance change $\Delta G(B)=G(B)-G(0)$ at
  different gate voltages at $T=400$ mK. The $n$th curve from bottom
  to top is shifted by $0.2\times (n-1)$ $\upmu$S vertically for
  clarity.}
\label{fig:GBVg}
\end{figure}

\begin{figure}[tbh]
\centering{\includegraphics[width=0.55\columnwidth]{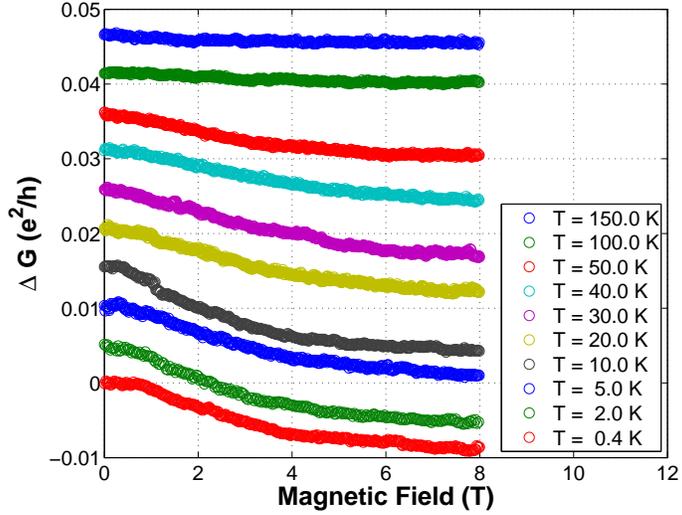}}
\caption{Magneto-conductance change $\Delta G(B)=G(B)-G(0)$ at
  different temperatures at $V_{g}=-4$ V. The $n$th curve from bottom
  to top is shifted by $0.2\times (n-1)$ $\upmu$S vertically for
  clarity.}
\label{fig:GBT}
\end{figure}

Magneto-conductance measurements were carried out with the external
magnetic field applied perpendicular to the axis of the nanowire as
well as the substrate (inset of Fig.~\ref{fig:GB}). The two-terminal
magneto-conductance $G$ of the nanowire was measured using quasi-dc
lock-in technique with a bias voltage of $40$ mV at $11.3$ Hz in a
Quantum Design Physical Properties Measurement System cryostat. Data
from one of the magneto-conductance measurements were plotted in
Fig.~\ref{fig:GB}, in which a clear magneto-conductance peak (blue
dotted line) at the center of magnetic field was observed.
Magneto-conductance data at different gate voltages and different
temperatures are plotted in Fig.~\ref{fig:GBVg} and Fig.~\ref{fig:GBT}
in the Supporting Information. It is well known that the quantum
interference of the electron wave functions reduces (increases)
backscattering of the electrons from impurities and therefore,
increases (decreases) the conductance from its Drude value leading to
weak antilocalization (localization) effects in systems with (without)
strong spin--orbit interactions \cite {Hikami}. Applying a magnetic
filed perpendicular to the sample destroys these interference effects
and restores the conductance to its Drude value. The present
observation is consistent with the suppression of the weak
antilocalization by the external magnetic field. The presence of weak
antilocalization in our experiment indicates strong spin--orbit
coupling in this system \cite{Roddaro2008}. It is important to note
that unlike previous reports on arrays of quantum dots
\cite{Faniel2007}, nanowires \cite{Kurdak} or two dimensional system
\cite{Koga2002}, the interpretation of our data is based on
weak-antilocalization effect of hole gas in an individual nanowire
without any average \cite{Hansen}.

In an one-dimensional system, at a magnetic field $B$, the weak
antilocalization correction including the spin--orbit interaction
effect to the conductance is given by \cite{Hikami, Beenakker, Kurdak,
  Hansen}:

\begin{eqnarray}
  G(B)=G_{0}-\frac{2e^{2}}{hL}\Bigg[ &&\frac{3}{2}\left( \frac{1}{l_{\phi }^{2}
    }+\frac{4}{3l_{SO}^{2}}+\frac{1}{D\tau _{B}(B)}\right) ^{-1/2}  \notag \\
  &&-\frac{3}{2}\left( \frac{1}{l_{\phi }^{2}}+\frac{4}{3l_{SO}^{2}}+\frac{1}{
      l_{e}^{2}}+\frac{1}{D\tau _{B}(B)}\right) ^{-1/2}  \notag \\
  &&-\frac{1}{2}\left( \frac{1}{l_{\phi }^{2}}+\frac{1}{D\tau _{B}(B)}\right)
  ^{-1/2}  \notag \\
  &&+\frac{1}{2}\left( \frac{1}{l_{\phi }^{2}}+\frac{1}{l_{e}^{2}}+\frac{1}{
      D\tau _{B}(B)}\right) ^{-1/2}\Bigg].  \label{eq:magneto-conductance}
\end{eqnarray}

\begin{figure}[tbph]
  \centering{\subfigure[gate voltage dependence]{\label{fig:Len_Vg}
      \includegraphics[width=0.45\columnwidth]{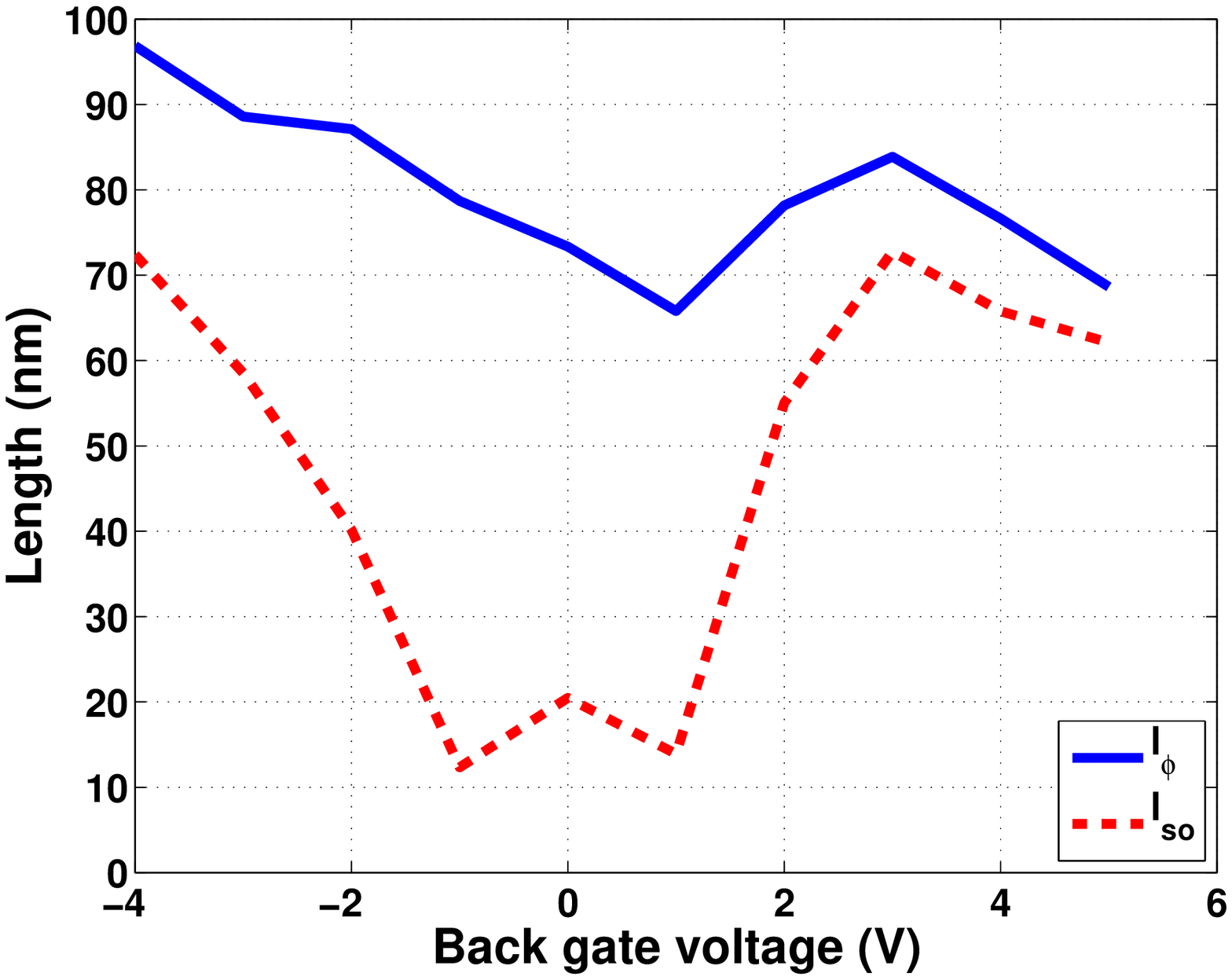}}
    \subfigure[temperature dependence]
    {\label{fig:Tau_T}\includegraphics[width=0.45\columnwidth]{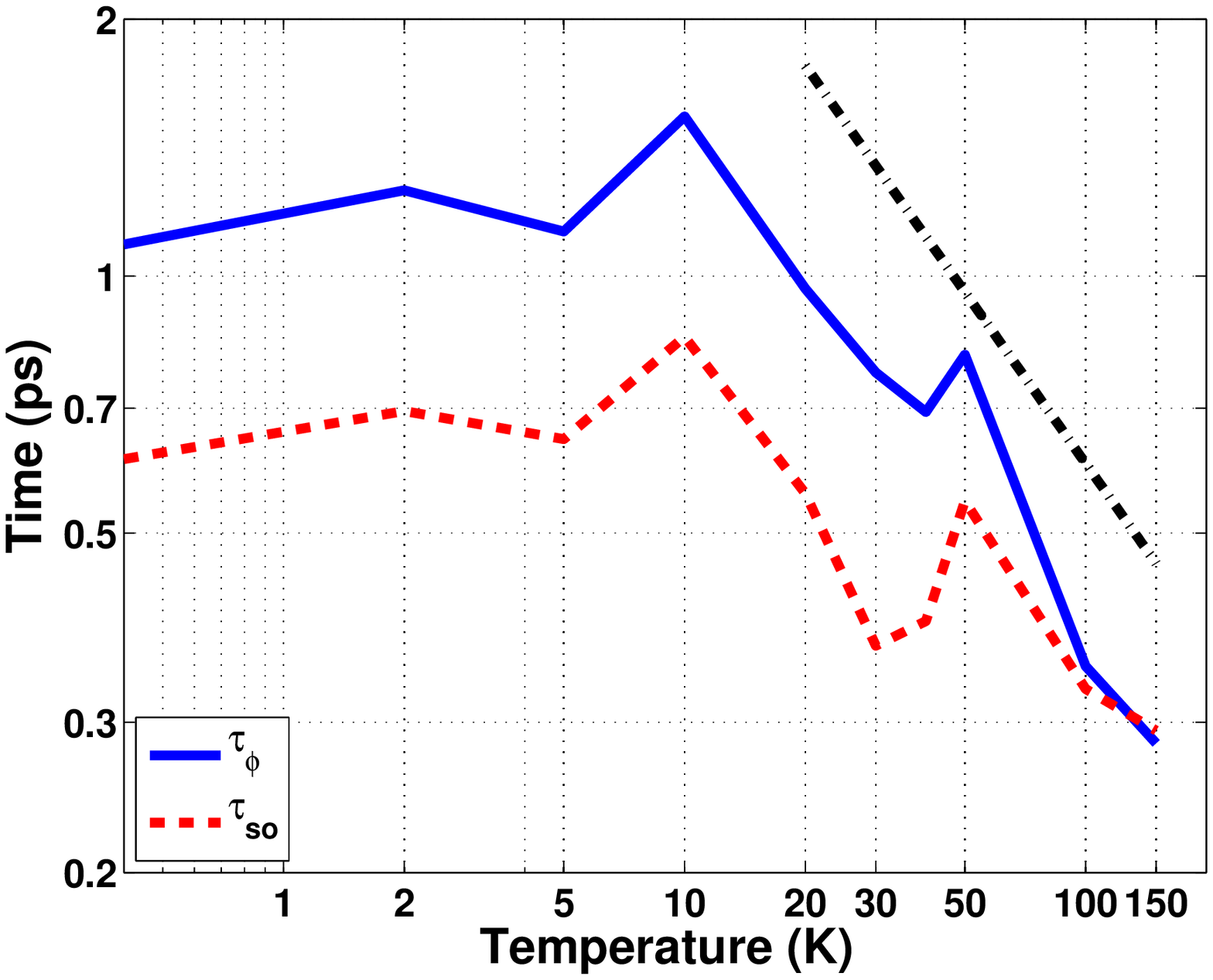}}
  }
  \caption{\Subref{fig:Len_Vg} The extracted phase coherence (blue
    solid line) and spin--orbit coherence length (red dashed line) as
    a function of back gate voltage at $ T=400$ mK. \Subref{fig:Tau_T}
    The extracted phase relaxation time (blue solid line) and
    spin-orbit relaxation time (red dashed line) as a function of
    temperature at $V_{g}=-4$ V. The black dash-dot line in panel
    \Subref{fig:Tau_T} indicates $\tau \propto T^{-2/3}$. }
\label{fig:Length_Tau}
\end{figure}

Here, $e$ is the electric charge of electron, $h$ is the Planck
constant, $l_{\phi }$ is phase coherence length, $l_{SO}$ is the
spin--orbit coherence length, $l_{e}=v_{F}\mu m^{*}/e$ is the elastic
mean-free path. $D=v_{F}l_{e}/2$ is the diffusion coefficient
\cite{Beenakker}. $G_{0}$ is the background conductance (Drude value)
without the localization or antilocalization correction. $\tau_{B}(B)
= 9.5l_{B}^{4}/(w^{3}v_{F}) + 4.8l_{e}l_{B}^{2}/(w^{2}v_{F})$ is the
relaxation time due to external magnetic field, for which we use the
expression in the case of specular boundary scattering in
one-dimensional channel \cite{Beenakker}. $w$ is the diameter of the
conductance channel (nanowire diameter). $m^{*}$ is the effective mass
of hole in Ge. The hole density $n_{d}$ and mobility $\mu$ can be
estimated from the transport data. $\lambda_{F} =
2(3n_{d}/\pi)^{-1/3}$ is the Fermi wavelength and
$v_{F}=2\pi/\lambda_{F}$ is the Fermi wave velocity. The estimated
$l_{e}$ value is around $22\sim 48$ nm, and the estimated
$\lambda_{F}$ value is around $5\sim 8.5$ nm (Both $l_{e}$ and
$\lambda_{F}$ depend on the gate voltage and the temperature). Since
our system is in the regime $L\gg l_{e}\gg w > \lambda_{F}$,
Eq.~\ref{eq:magneto-conductance} is suitable for our case. In
addition, since this fitting equation is only valid for $l_{B}=(\hbar
/eB)^{1/2}>w$, we restricted our fitting to the data where the
magnetic field is smaller than $6$ T. Overall there are three
parameters $l_{\phi }$ , $l_{SO}$ and $G_{0}$ in
Eq.~\ref{eq:magneto-conductance} that are used to fit our experimental
data. One of the fitting curves is indicated by the red dashed line in
Fig.~\ref{fig:GB} and the fitted $l_{\phi }$ at $400$ mK are shown in
Fig.~\ref{fig:Len_Vg} for different back gate voltages $V_{g}$. There
is an apparent overall decrease of $l_{\phi }$ in the range between
$V_{g}=-4\sim 5$ V. This can be explained as the holes lose phase
coherence as the device becoming more insulated \cite{Nakamura2009}.
This change demonstrates that the carrier coherence properties can be
controlled by tuning the gate voltage over a wide range
\cite{Hansen,Dhara2009}. From the fitting data, we also obtained the
phase relaxation time $\tau _{\phi }=l_{\phi }^{2}/D$, which is
consistent with the values obtained from weak localization of holes in
\textit{p}-SiGe quantum wells \cite {Senz2000,Coleridge2002}. The
extracted phase coherence times at the same gate voltage follows a
power law dependence of temperature $\tau _{\phi }\propto T^{-2/3}$,
as shown in Fig.~\ref{fig:Tau_T}, which is observed previously in
several other one-dimensional systems
\cite{Altshuler1982,Lin1987,Pierre2003,Chiquito2007,Gao2009}. The
$T^{-2/3}$ dependence of $\tau _{\phi }$ indicates that the scattering
mechanism is dominated by the Nyquist process, in which the inelastic
hole--hole collision happens with small energy exchange
\cite{Altshuler1982}. At low temperatures, $\tau _{\phi }$ is found to
saturate, which is a signature of either the heating the carriers by
current \cite{RueB2007} or external microwave noise \cite{Khavin1998}.

Fig.~\ref{fig:Len_Vg} also shows the extracted spin--orbit coherence
length $l_{SO}$ as a function of the gate voltages. The extracted
values of $l_{SO}$ are in the range $70\sim 80$ nm at $400$ mK. The
spin--orbit relaxation time shown in Fig.~\ref{fig:Tau_T} is deduced
from $l_{SO}=(D\tau_{SO})^{1/2}$. It is interesting to note that
unlike earlier studies on electron gas systems which show that
$\tau_{SO}$ is independent of temperature \cite{Hansen,Guzenko2007},
in our case we found $\tau_{SO}$ decreases by about two folds as the
temperature is increased to $150$ K from $10$ K, indicating that
hole--hole scattering and phonon scattering accelerate the spin
relaxation at high temperatures \cite{Kallaher2010}.

\begin{figure}[tbph]
  \centering{ \subfigure[density dependence]{\label{fig:G_nd}
      \includegraphics[width=0.45\columnwidth]{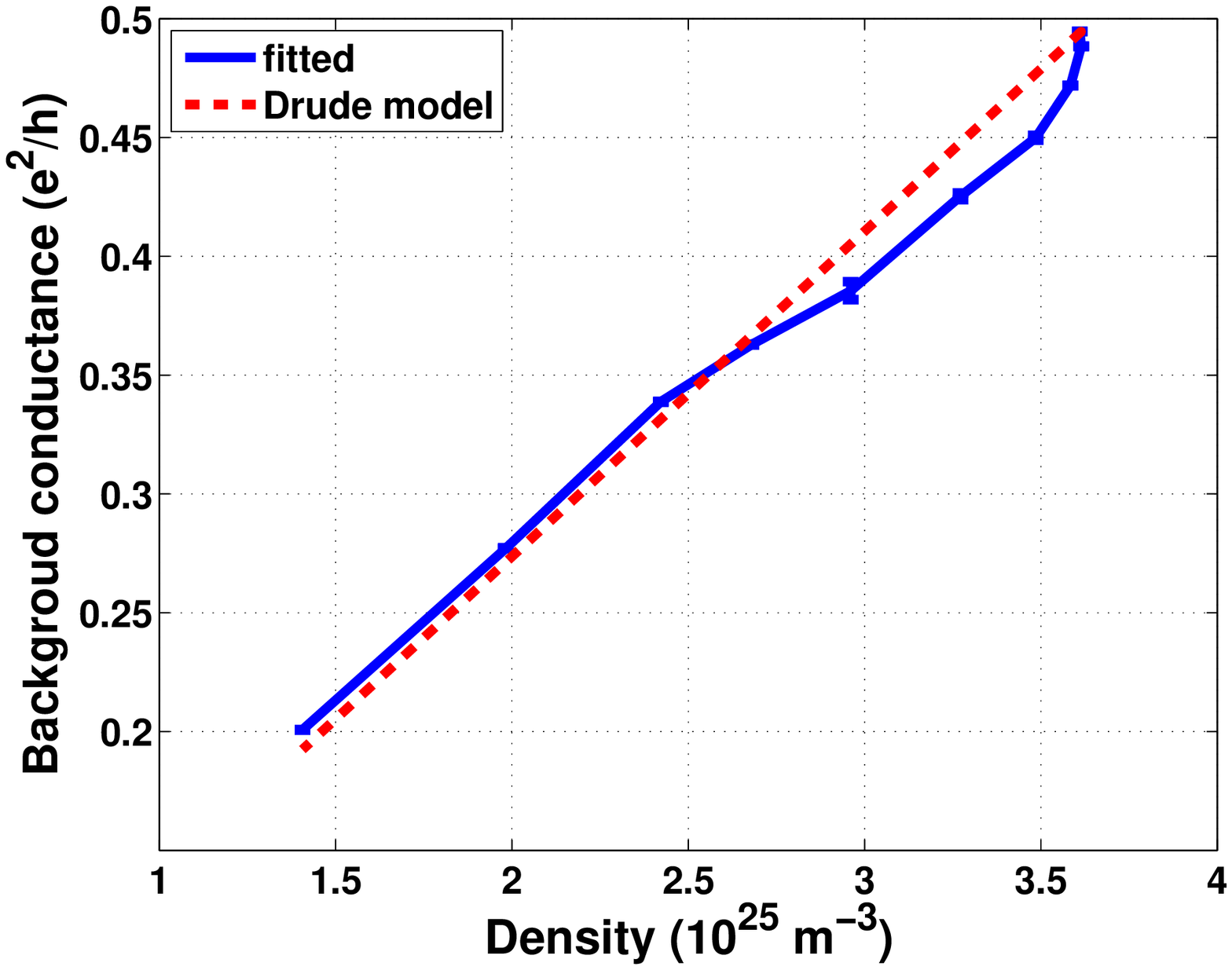}}
      \subfigure[temperature dependence]
      {\label{fig:G_T}\includegraphics[width=0.45\columnwidth]{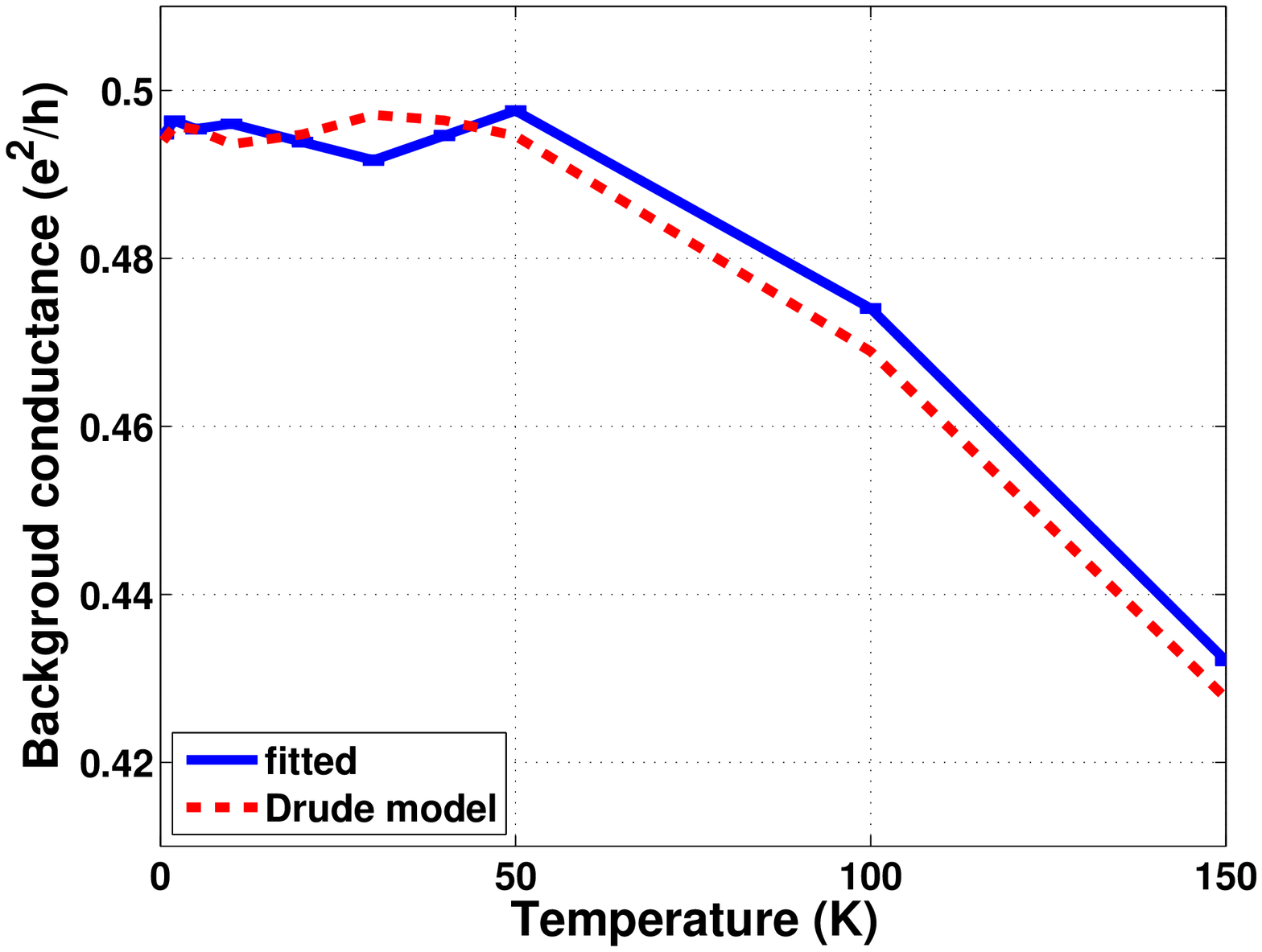}}}
    \caption{The background conductances $G_{0}$ obtained from fitting
      (blue solid line, shown with error bar) and from the Drude model
      (red dashed line) as a function of \Subref{fig:G_nd} hole
      density and \Subref{fig:G_T} temperature.}
\label{fig:G}
\end{figure}

To further confirm the validity of the fitting method, the background
conductance $G_{0}$ obtained from fitting using Eq.~\ref
{eq:magneto-conductance} was compared with that estimated using the
Drude classical conductance model for the one-dimensional nanowire:

\begin{equation}
  G_{Drude}=\frac{\pi w^{2}}{4L}n_{d}e\mu,  \label{eq:drude}
\end{equation}
where $n_{d}$ and $\mu$ are density and mobility of the nanowire
estimated from transport data. We plotted the fitted background
conductance as well as the values estimated from the Drude model in
Fig.~\ref{fig:G} for different densities (Fig.~\ref{fig:G_nd}) and
different temperatures (Fig.~\ref{fig:G_T}). It is found the extracted
$G_{0}$ values are in good agreement with the estimated Drude
conductance, which implies the weak antilocalization dominates the
transport and the model is well described by
Eq.~(\ref{eq:magneto-conductance}). Fig.~\ref{fig:G_nd} shows that the
conductance decreases monotonically with the decrease of the hole
density, as expected. We also found in Fig.~\ref{fig:G_T} that the
background conductance keeps increasing when the temperature decreases
before $T>30$ K, which indicates the suppression of hole-phonon
interaction at lower temperatures. In addition, $G_{0}$ saturates
below $30$ K, which is likely due to the effect of residual impurity
scattering inside the nanowire \cite{Dhara2009}.

\begin{figure}[tbh]
\centering{\includegraphics[width=0.55\columnwidth]{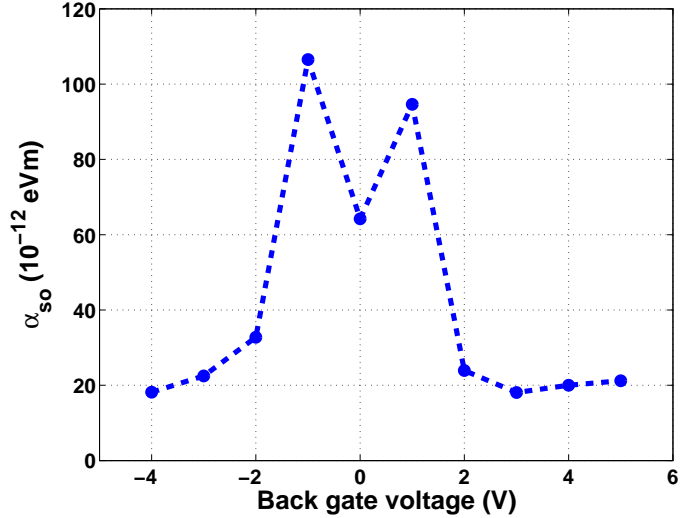}}
\caption{Plot of spin--orbit coupling strength $\alpha_{SO}$ as the
  function of the back gate voltage.}
\label{fig:alpha}
\end{figure}

We now discuss spin--orbit coupling in this system. Since both bulk Si
and Ge have an inversion center \cite{Golub2004}, in our analysis we
do not include the Dresselhaus effect, which is induced by bulk
inversion asymmetry \cite{Dresselhaus1955}. The Elliott-Yafet effect,
which is caused by the lattice vibrations and impurity scattering
\cite{Elliott1954}, was also neglected due to the high hole mobility
and the observed effect that the spin--orbit coherence length $l_{SO}$
is a strong function of $V_{g}$ (Fig.~\ref{fig:Len_Vg}) but the
mobility $\mu$ is insensitive to $V_{g}$. To obtain the spin--orbit
coupling strength $\alpha _{SO}$, we only include the Rashba effect,
which originates from the structural inversion asymmetry of the device
geometry and can me modulated by external potential \cite
{Rashba1984,Koga2002,Tahan2005}:

\begin{equation}
  \alpha _{SO}=\frac{\hbar}{k_{F}\sqrt{2\tau_{e}\tau_{SO}}},  \label{eq:spin-orbit}
\end{equation}
where $k_{F}$ is the Fermi wave number, and $\tau_{e}$ is the
transport relaxation time. In this way, the spin--orbit coupling
strength $\alpha _{SO}$ can be deduced from the weak antilocalization
measurements, as shown in Fig.~\ref{fig:alpha}. Significantly, we note
that the spin--orbit coupling strength decreases by more than five
folds as the absolute value of the back gate voltage $V_{g}$
increases. The gate tunable spin--orbit coupling in Ge/Si core/shell
nanowires in turn suggests the Ge/Si nanowire system is a suitable
candidate in applications of spintronics device \cite{Datta1990APL}.
In addition, a nonmonotonic back gate voltage dependence was observed
for $l_{\phi}$, $l_{SO}$ (Fig.~\ref{fig:Len_Vg}) and $\alpha _{SO}$
(Fig.~\ref{fig:alpha}) around $V_{g}=0$ V \cite{Dhara2009}. The
nonmonotonic behavior might be caused by intersubband scattering
between multiple subbands involved into the magneto-transport
\cite{Lu2005,Gao2009}, or how the original asymmetry of the
confinement well profile is canceled by external electric field at
small field \cite{Tahan2005}, which need further investigation.

In conclusion, we observed weak antilocalization of one-dimensional
holes in an individual Ge/Si core/shell nanowires with strong
spin--orbit coupling. The phase coherence length, the spin--orbit
coherence length as well as the spin--orbit coupling strength and
background conductance were extracted by fitting the experimental data
at different temperatures and gate voltages. Both the phase coherence
length and the spin--orbit coupling strength were found to be
adjustable by an external electric field. Our results illustrate the
potential of chemically synthesized Ge/Si core/shell nanowires in
future spintronics applications, and using this system as a platform
for studying coherent spintronics phenomena of holes in
low-dimensions.

\section*{Acknowledgment.}
This work was supported by the National Science Foundation
(ECS-0601478), the National Basic Research Program of China (Grants
No. 2009CB929600, No. 2006CB921900), and the National Natural Science
Foundation of China (Grants No. 10804104, No. 10874163, No. 10934006).

\section*{Supporting Information Available.}
Conductance and magneto-conductance at different temperatures and gate
biases and effects of Pd electrodes on the magneto-conductance
effects. This material is available free of charge via the Internet at
http://pubs.acs.org.

%%%%%%%%%%%%%%%%%%%%%%%%%%%%%%%%%%%%%%%%%%%%%%%%%%%%%%%%%%%%%%%%%%%%%%%%%%%%%%%%%%

\section*{References and Notes}

\section*{Supporting Information}

\newpage

\textbf{Discussion on the possible influence of Pd contacts on the
  observed magnetoresistance effects.}

We believe the observed magnetoresistance effects are intrinsic to the
Ge/Si nanowire system due to the following reasons:

From the data in Phys. Rev. B 39, 915 (1989) and Phys. Rev. B 39, 3015
(1989), we resistance of the Pd leads was estimated to be less than
$5$ k$\Upomega$ ($L\times W\times H = 100$ $\upmu$m $\times 400$ nm
$\times 50$ nm), which is much smaller than the resistance of the
nanowire ($50\sim 100$ k$\Upomega$). In addition, the
magnetoconductance effect in Pd films has been found to be only around
$0.5\%$ based on the data in the references listed above. This effect
is too small to explain the observed magnetoconductance of around
$3\%$ observed in our system.

Furthermore, the magnetoconductance effects induced by the Pd contacts
will be relevant when the magnetic length
$l_{B}=\sqrt{\frac{\hbar}{eB}}>w=50$ nm (thickness of the Pd
electrodes), i.e. for magnetic field $\left\vert B\right\vert <0.26$
T. However, the magnetoconductance peak observed here covers several
Tesla.

Finally, Aluminum contacted Ge/Si core/shell nanowire devices have
also been tested and similar magnetoconductance effects were observed
which are consistent with the results in this manuscript.

\end{document}